# Advances in ion-doping of Ca-Mg silicate bioceramics for bone tissue engineering

[*1]Ashkan Namdar, Erfan Salahinejad

Faculty of Materials Science and Engineering, K. N. Toosi University of Technology, Tehran, Iran

## Abstract

The use of bioceramics as hard tissue substitutes is extensive due to their excellent biocompatible and osteogenic behaviors. Among various bioceramics, Ca-Mg silicates are unique from the viewpoints of osteoinductive and mechanical properties, as well as their outstanding osteoconductive and angiogenic behaviors owing to the release of Si, Ca and Mg. Despite these distinct advantages, different compositions of these bioceramics still require mechanical and biological enhancements for further applications. For this purpose, doping with some ions like $F^-$, $Sr^{2+}$, $Cu^{2+}$, $Eu^{2+}$, $Ba^+$, $Ce^{3+}$ and some alkali cations has been proved to be a valued approach. This review attempts to bring together areas for the performance improvement of the further researched Ca-Mg silicates (i.e., diopside, akermanite, bredigite and monticellite) and the alteration of these compositions via ion-doping. It is concluded that a correct choice of dopants incorporated at the optimal concentration makes these silicates ideal bone substitutes competing or even superior to calcium phosphates (apatites) and bioglasses which are known as the most prominent bioceramics.

[1*] Corresponding Author: Email Address: <salahinejad@kntu.ac.ir>

**Table of Content**

# 1. Introduction

Today's world, reaching a summit of scientific and technological improvements, is observing a severe number of accidents by which bones can be damaged critically. Besides, by enhancing life habits, the elderly live longer and as a result, osteoporosis and other possible bone defects occur. Diseases like bone cancer and their treatments also excessively intensify the possibility of bone flaws [1,2]. According to GlobeNewswire [3], a total count of nearly 22.3 million orthopedic surgeries was undergone in 2017, and a compound annular growth rate of 3.8% is estimated in this regard [4]. All of the aforementioned issues indicate that developing materials well-suited to heal fractured or damaged bones is of essential significance in biomedicine.

For effective bone regeneration, there are four significant factors to be taken into account, including biocompatibility, biodegradability, mechanical properties and pore conditions [5]. A biocompatible bone substitute does not induce any toxic behaviors, severe localized or systematic side effects, immunity system breakage, mutations and carcinogenic reactions. Also, the ability of a bone-substituting implant to be resorbed in the body can introduce a variety of new features, eliminating the need to a secondary operation. For an optimal biodegradable bone substitute, the resorption rate should be lower than the bone regeneration rate. The substitute's mechanical properties should suffice and are close to those of the target cortical or cancellous bone (Table 1). Indeed, pore conditions of the substitute (pore size and distribution) determine some properties like mechanical behavior, bioactivity, biodegradability, vascularization, cell attachment and proliferation [6]. It is well-established that the optimal pore size for vascularization is between 100 and 400 μm [7], whereas it is 20 to 1500 μm for proper cell attachment and proliferation [8].

Table 1. A summary of mechanical properties of human bones [9–11]

| Properties | Cortical bone | Cancellous bone |
|---|---|---|
| Tensile strength (MPa) | 50-150 | 10-100 |
| Compressive strength (MPa) | 130-230 | 2-12 |
| Shear strength (MPa) | 53-70 | 53-70 |
| Elastic modulus (GPa) | 7-30 | 0.02-0.5 |
| Fracture toughness (MPa.m$^{1/2}$) | 2.4-6.7 | 1.6 |

Bioceramics are one of the most appropriate choices for biomedicine. They can harmlessly be used in the human body as they are biocompatible. Some bioceramic variations exhibit bioactivity since they offer a suitable attachment to bone tissue and human flesh. Osteoconduction is another well-established property of bioceramics. Calcium phosphates (for example, tricalcium phosphate and hydroxyapatite) are typical examples of valid bioceramics widely used in biomedicine due to their excellent biocompatibility and osteoconductivity [12,13]. In addition, biodegradability is a favorable property when using bioceramics, regarding how they lessen the possibility of secondary operations. For biodegradable bioceramics, the release of incorporated ions into the body fluid and their interconnection give rise to the precipitation of mainly carbonated apatite layers [14]. From a mechanical point of view, elasticity, strength and fatigue behavior are the most significant features of bone substitutes. It is well-established that the elastic modulus of a bone graft should be near to that of the target area bone [15]. Even though calcium phosphates generally exhibit excellent biocompatibility and bioactivity, they suffer from inadequate mechanical properties. Moreover, newer generations of bioceramics, e.g., bioactive glasses and calcium silicates, do not show an outstanding mechanical behavior, typically low fracture toughness [16,17]. Using modified calcium silicates and calcium-magnesium silicates could be a backdoor to have both mechanical and biological excellencies in performance.

Calcium-containing silicate bioceramics, in comparison to most calcium phosphates, can further promote cell differentiation and proliferation [18–20]. It is mainly owing to the ionic release of Si and Ca which are well-known as osteoblast promoters. Results obtained by Xynos et al. [21] designate the probability of bone growth factor transcription through these ions. Calcium-magnesium silicates are a distinct group of the aforementioned silicates by which bone regeneration is further improved due to the positive impact of Mg [19,22]. The dissolution rate of Ca-Mg silicates is moderate compared to other Ca silicates and glasses owing to the replacement of Ca-O bonds by stabler Mg-O bonds. The same reason contributes to their priority in mechanical enhancement and higher crystallinity [23]. Akermanite ($Ca_2MgSi_2O_7$), diopside ($CaMgSi_2O_6$), bredigite ($Ca_7MgSi_4O_{16}$) and monticellite ($CaMgSiO_4$) are the most frequently researched Ca-Mg silicates in recent years. All of the mentioned compositions possess suitable biocompatibility and bioactivity [24,25].

Despite the priority of Ca-Mg silicates in mechanical behavior over glasses, there are still some areas for the further improvement of these bioceramics. One approach to altering the bio-performance of bioceramics is the use of ion doping. Mainly researched doping agents in Ca-Mg silicates are $F^-$, $Sr^{2+}$, $Cu^{2+}$, $Eu^{2+}$, $Ba^+$, $Ce^{2+}$ and alkali metal cations. For instance, diopside possesses considerable mechanical properties, but it lacks sufficient bioactivity [19], which can be effectively addressed by $Sr^{2+}$, $K^+$, $Li^+$ and $F^-$ doping. Similar to other biomaterials, $Cu^{2+}$ incorporated into Ca-Mg silicates inhibits bacteria growth [26], where most of the Ca-Mg silicates suffer from insignificant antibacterial properties. Despite adequate biological properties, mechanical properties of akermanite need to be improved, which can be achieved by $Ba^{2+}$ incorporation [27]. Rare earth elements (REEs) doping in Mg-Ca silicates has been also indicated to enhance apatite formability and introduce photoluminescence [28,29]. All of these statements represent that a proper dopant at optimal levels can overcome the drawbacks of Ca-Mg silicates, enhance their present properties or induce new features, in terms of mechanical properties, osteogenesis, angiogenesis, photoluminescence, antibacterial properties, etc.

Despite the publication of several review papers on doping of other bioceramics, including apatites [2,12] and bioglasses [13,14,30], no review has been published on doped Ca-Mg biosilicates, to the best of our knowledge. The issue of doping has been dedicatedly addressed in none of review papers published on Ca-Mg biosilicates [31,32], whereas a good number of research papers signify the considerable influence of doping on the mechanical and biological performance of Ca-Mg biosilicates. Accordingly, this paper reviews advances in doping of Ca-Mg silicates used in biomedical applications.

## 2. Role of Mg, Ca and Si

Speaking of Ca-Mg silicates, a proper understanding of the contribution of each main constituent of them is of essential significance. In this regard, this chapter describes the biological and mechanical attributions of Mg, Ca and Si ions, as tabulated in Table 2.

Table 2. Roles of the major constituents of Ca-Mg silicates

| Element | Role |
| --- | --- |
| **Silicon** | Enhances apatite-formation ability and osteoinduction |
| **Magnesium** | Effective in over 300 enzymic reactions, RNA and DNA stabilizer, improves osteoinduction and bioactivity, retards the degradation and apatite-formation ability kinetics, improves mechanical properties |
| **Calcium** | Most major inorganic constitute of bone, excites osteogenesis, angiogenesis and IGF |

Silicon is the essential constituent of Ca-Mg silicates. Its first significance is in the apatite formation mechanism through the surface production of a silanol gel film. A five-step procedure has been proposed for apatite formation [33]; (1) Si-O-M (M, an alkali-earth metal) modified networks are rapidly transformed to $(Si-O-H)^+$ groups via the ion exchange of $M^{2+}$ and $H^+$, (2) the evolution of the first step increases physiological pH due to the consumption of $H^+$ accompanied by surface hydroxyl generation. $OH^-$ causes tears of Si-O-Si bonds to create silanol groups, (3) the produced silanol (SiOH)-rich gel layer with a surface negative charge attracts $Ca^{2+}$ from bones, blood vessels and Ca-Mg silicate, (4) owing to the presence of the newly formed positively-charged layer of Ca, $PO_4^{3-}$ migrates toward the surface. Steps 3 and 4 take place orderly and lastly, (5) the amorphous structure of the Ca/P layer, known as apatite, is crystallized [33]. Si also presents osteoinductivity, which is responsible for the osteoinductive behavior of bioglasses, for instance [34].

Calcium is the most essential alkali-earth element in the human body, especially in teeth and bones, as 98% of its total amount is concentrated in these tissues. It excites osteo/angiogenesis favored for differentiation at lower concentrations. Medium concentrations of calcium help the mineralization of extracellular matrices (ECMs). However, excessive amounts of this species are toxic. Calcium can improve the usefulness of insulin-like growth factors (IGFs), an essential growth factor in the osteoblastic mechanism [19]. The addition of Ca to silicate lowers the connectivity of the silicate network, resulting in a higher degradation rate of the bioceramics. This is indeed the first reason for the bioactivity of Ca-containing silicates [35].

From the biological point of view, Mg behaves as a precursor for over 300 enzymic reactions and DNA and RNA stabilizer [36]. Mg is also an essential element in bone generation via escalating osteoblast activity, where a lack of Mg can lead to osteoporosis and bone brittleness [37]. On the contrary, Mg lowers the dissolution rate of silicates (owing to the more covalent character of the Mg-O bond than the Ca-O one) [38,39], the nucleation rate of apatite precipitation and the crystallinity of deposited apatite [40].

## 3. Weaknesses and areas for improvement of medically-researched doped Ca-Mg silicates

Prior to the introduction of Ca-Mg silicates, Ca silicates were synthesized as a derivative of Hench et al. [41]'s bioglass. Ca silicates express remarkable bioactivity and degradability, but a possibility of toxicity is imminent at high dissolution rates. Moreover, weak mechanical properties make them an undesirable choice for medical purposes, unless modifications take place [24]. In the case of ion introduction, $Mg^{2+}$ can be added to Ca silicates in the form of MgO, giving Ca-Mg silicates. Despite this addition, there are still drawbacks in practice for Ca-Mg silicate systems, that might be enhanced through ion doping. Besides, some other areas for development are available for Ca-Mg silicates. This chapter draws out these shortcomings and possibilities to be improved, as summarized in Table 3. Figure 1 also illustrates the structure of diopside, akermanite, bredigite and monticellite as Ca-Mg silicates that have been studied in the case of medical doping.

Table 3. A comparison of four medical doped Ca-Mg silicates

| Silicate name | Chemical formula | Pros | Cons |
|---|---|---|---|
| Diopside | $CaMgSi_2O_6$ | Great mechanical strength | Worst bioactivity |
| Akermanite | $Ca_2MgSi_2O_7$ | Great osteogenesis facilitation | Inadequate mechanical properties |
| **Bredigite** | $Ca_7MgSi_4O_{16}$ | Highest apatite-formation ability | Worse mechanical properties and biocompatibility |
| **Monticellite** | $CaMgSiO_4$ | Suitable mechanical properties | Inadequate bioactivity |

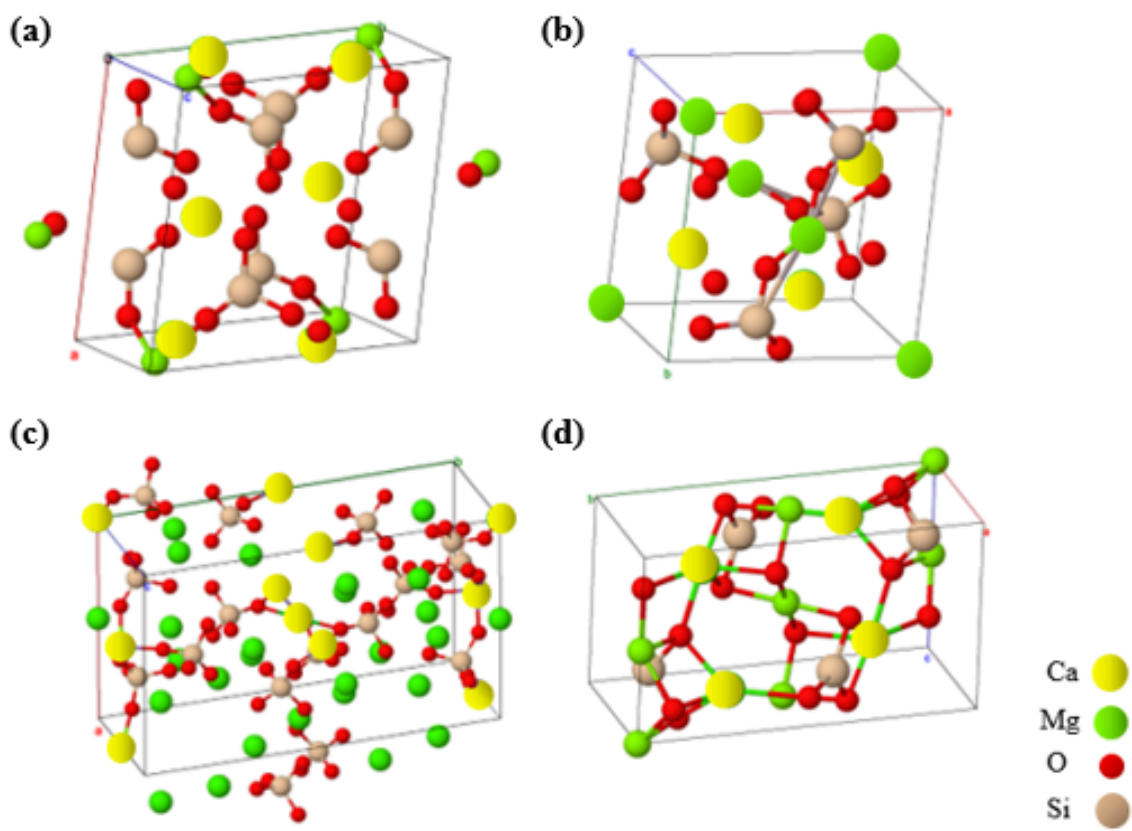

Figure 1. Ionic arrangement of diopside (a), akermanite (b), bredigite (c) and monticellite (d).

It is noteworthy that even though the bioactivity of Ca-Mg silicates differs from each other, they can be better than HA and most bioactive glasses [24,42–44]. Interestingly, as the CaO/MgO ratio increases, better bioactivity results. On the contrary, at lower values of the CaO/MgO ratio, higher mechanical properties are revealed. This is in agreement with the fact that $Mg^{2+}$ lowers the dissolution rate to improve ceramics mechanically. Additionally, by increasing the $Ca^{2+}$ content, dissolution becomes more rapid; thus, better bioactivity (concerning apatite formation) is expected.

### 3.1. Diopside

Diopside possesses outstanding mechanical strength compared to the other participants of Ca-Mg silicates. It has an incredible fracture toughness of almost 3.5 $MPa.m^{1/2}$, bending strength up to 300 MPa and elastic modulus of about 170 GPa [45]. In a study by Wu et al. [46], the compressive strength of porous diopside with 75-80% porosity was similar to that of spongy bone with 70-90% porosity (0.63-1.36 and 0.20-4.00 MPa, respectively). The one major shortcoming of diopside is its low biodegradability and bioactivity [31,32]. Its high Si content, ensuring high bridging oxygen levels and thereby low dissolution rate, is in charge of such behavior. Considering these facts, a priority of biodegradation and bioactivity enhancements is the most significant area to heighten the desirability of diopside.

### 3.2. Akermanite

Akermanite has been proven to have significant bioactivity and osteogenesis behaviors [47,48]. Its mechanical properties are not ideal, with a fracture toughness of nearly 1.8 $MPa.m^{1/2}$, Young's modulus of about 42 GPa and bending strength of almost 176 MPa [49,50]. It is worth noting that these numbers might change owing to different sintering variables. Conclusively, akermanite possesses both suitable bioactivity and adequate mechanical strength, but low fracture toughness might become a pickle in biomedical uses.

### *3.3. Bredigite*

Bredigite shows even worse mechanical behaviors than akermanite, with a fracture toughness of almost 1.6 MPa.m$^{1/2}$ [51]. Its biodegradation rate is near to that of β-TCP, which is higher than akermanite. The higher CaO/MgO ratio and lower $SiO_2$ content of bredigite are responsible for its higher degradation rate. Accordingly, a higher apatite-formation ability has been reported for bredigite compared to diopside and akermanite [19]. However, the high degradability of bredigite causes the metabolic alkalosis effect, deteriorating biocompatibility and overall bioactivity governed by cell activity [51].

### *3.4. Monticellite*

Despite the same crystalline structure of monticellite and bredigite, the CaO/MgO ratio and behaviors of monticellite and diopside are similar in terms of low apatite-formation ability and high mechanical properties [32]. Comparatively, monticellite presents slightly higher bioactivity than diopside, but lower than the two other aforementioned silicates [19,31].

## 4. Synthesis approaches of doped Ca-Mg silicates

There are three major processes to synthesize doped Ca-Mg silicates, including milling, coprecipitation and sol-gel methods (Figure 2), in which dopant compounds are used as one of the precursors. In the milling method, principal precursors and dopant oxides are added to a milling machine to be ground together. This process leads to the fragment of the loaded particles and the formation of virgin surfaces which tend to bond with each other. The short relaxation time of about $10^{-7}$-$10^{-3}$ sec, enhanced driving force and reduced activation energy during milling enhance the diffusivity of the components and facilitate atomic/ionic intermixing, resulting in the designed synthesis [52–54]. Afterwards, the power can be compressed and sintered at predetermined temperatures to obtain bulk substances. Achieving submicron sizes and uniformity in the particle size without using extra additives is of the main advantages of this route. Nonetheless, long processing time, uncertainty in reaching the desired composition and unavailability to obtain nanosized powders are the most prominent drawbacks of this process [55]. Concerning doped Ca-Mg silicates, Eu-, Sr-and Cu-doped akermanite samples were successfully synthesized by ball-milling of Mg, Ca, Si and dopant oxides and calcination at 900-1200 ºC [56–58].

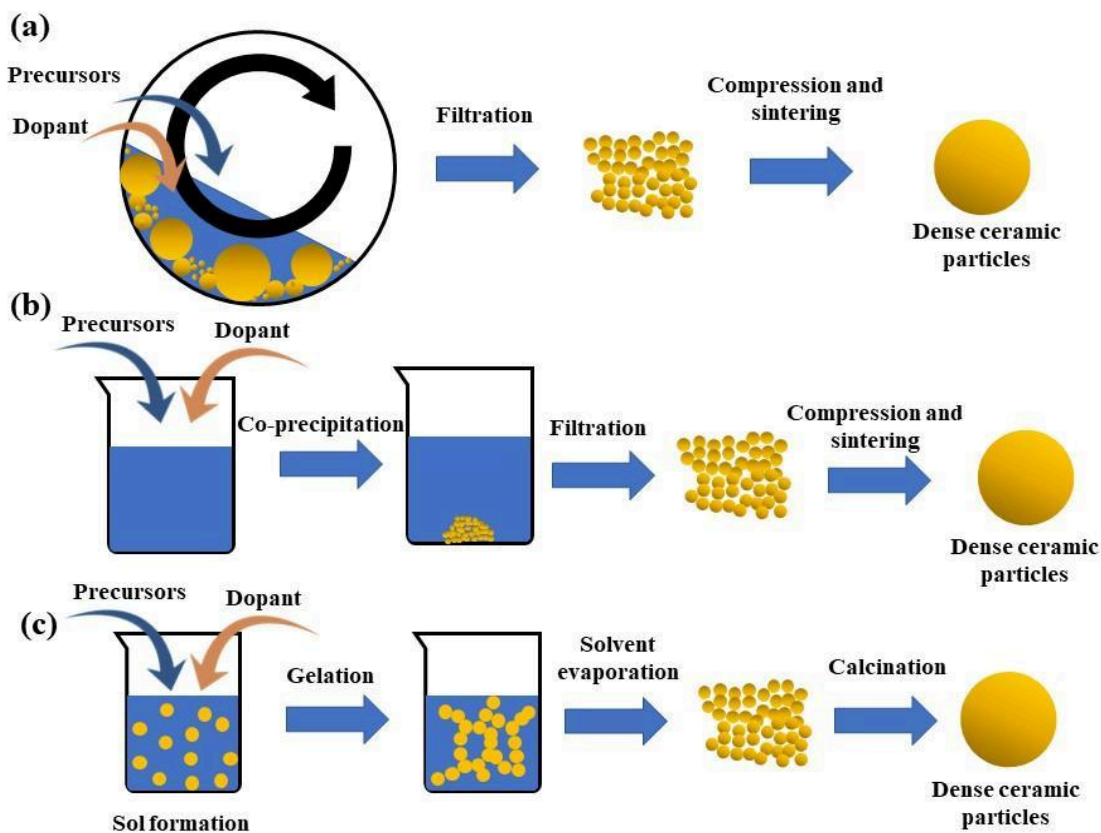

Figure 2. Illustration of milling (a), co-precipitation (b) and sol-gel (c) methods to fabricate ion-doped Ca-Mg silicates.

Co-precipitation involves the addition of precursors and dopant compounds to a solvent, followed by the addition of a precipitant to induce supersaturation, thereby forming precipitates. In this method, a large number of nuclei are created under the imposed supersaturation conditions. As the growth step, subsequent substances in the solvent are adsorbed on the nuclei or diffuse into the structure of the nuclei. Finally, particle coarsening (Ostwald ripening) and agglomeration occur, which play a crucial role in the final size, shape and morphology of the products [59–61]. The products are then processed by filtration or solvent evaporation and calcination at predetermined temperatures. This process is capable of producing micron and nanosized particles, whereas the particles are not necessarily of a uniform size and/or composition. Also, the morphological features of the particles are difficult to be controlled in this procedure [62]. Typically, Keihan et al. [63] and Su et al. [28] employed the inorganic-salt coprecipitation method to synthesize bredigite and akermanite, respectively, in which tetraethyl orthosilicate (TEOS), dopant and precursors chlorides were solved in ethanol. Then, an ammonia solution was added to form precipitates, followed by drying and calcination to obtain desired phases.

The sol-gel method is another wet-chemical synthesis process, in which a solution of precursors and dopant compounds is prepared and then gelled. First, a sol (colloidal suspension or solution) of metalloids or metals bound to proper ligands is formed. Then, the sol containing metal-ligand complexes undergoes hydrolysis or alcoholysis, followed by polycondensation and thereby the formation of macromolecules in a fractal polyfunctional design [60,64,65]. Subsequently, the evaporation of the solvent leaves a xerogel powder which is then calcinated to provide dense ceramic structures [66]. Chemical homogeneity, purity, narrow size distribution, uniformity and controllability over size and morphology are the major advantages of the sol-gel process, whereas harder control over porosity and high cost are its main disadvantages [62]. Regarding doped Ca-Mg silicates, a TEOS solution and hydrated nitrides of the metallic precursors (Mg and Ca) and dopant (Mg) were loaded in the akermanite stoichiometry and successfully deposited on Ti-6Al-4V substrates [37].

## 5. Dopants contributing to Ca-Mg silicate systems

In this chapter, the role of each dopant on the bio-performance of Ca-Mg silicates is reviewed from research works tabulated in Table 4.

Table 4. A summary of reviewed dopant systems and their results

| Matrix Ca-Mg silicate | Dopant | Production method | Concentrations | Results | Ref. |
|---|---|---|---|---|---|
| Diopside | $F^-$ and $Sr^{2+}$ | Co-precipitation | 0 mol% $Sr^{2+}$ 1 mol% $F^-$ | Limited Si release; improved apatite-formation ability; imposed toxicity | [67] |
| | | | 2 mol% $Sr^{2+}$ 0 mol% $F^-$ | Improved cell activity and cytocompatibility; lowered apatite formability | |
| | | | 2 mol% $Sr^{2+}$ 1 mol% $F^-$ | A combination of acceptable bioactivity and cytocompatibility | |
| | $Sr^{2+}$ | Co-precipitation | 2 mol% | Improved cell activity and cytocompatibility | [68] |
| | $Li^+$ | Co-precipitation | 2 mol% | Improved apatite formability; lowered cell viability | [69] |
| | $Na^+$ | | 2 mol% | Lowered apatite formability; improved cell viability | |
| | $K^+$ | | 2 mol% | A combination of acceptable apatite formability and cytocompatibility | |
| | $F^-$ | Co-precipitation | 1 mol% | Acceptable apatite formability | [70] |
| | | | 2 mol% | Acceptable apatite formability; cautions of toxicity | |
| | | | 5 mol% | Acceptable apatite formability; cautions of toxicity | |
| | | | 10 mol% | Lowered bioactivity; cytocompatibility cautions | |
| | | | 20 mol% | Lowered apatite formability | |
| | $F^-$ | Co-precipitation | 1 mol% | Improved mechanical properties and apatite formability | [71] |
| | $Ce^{3+}$ | Co-precipitation | 0.25 mol% | Improved apatite formability and mechanical properties | [28] |
| | | | 0.50 mol% | Lowered apatite formability; deteriorated mechanical properties; | |

| | | | 0.75 mol% | Lowered apatite formability | |
|---|---|---|---|---|---|
| | | | 1.00 mol% | Not notable bioactivity; highly improved mechanical properties | |
| **Bredigite** | $Cu^{2+}$ | Co-precipitation | 5 mol% | High level of vascularization; highly lowered cell activity | [26] |
| | $F^-$ | Co-precipitation | 0.5 wt% | Improved apatite formability; imposed a limited level of toxicity | [72] |
| | | | 1 wt% | A combination of outstanding apatite formability and mechanical strength | |
| **Akermanite** | $Sr^{2+}$ | Milling and calcination | 0.05 mol% | Acceptable apatite formability; improved cell activity and cytocompatibility | [57] |
| | | | 0.10 mol% | Improved release of Mg; Improved bioactivity | |
| | | | 0.15 mol% | Improved bioactivity; Improved cytocompatibility | |
| | $Cu^{2+}$ | Milling and calcination (mechanical activation) | 2.5 mol% | Improved mechanical properties and vascularization | [58] |
| | | | 3.5 mol% | Improved mechanical properties and vascularization | |
| | | | 4.5 mol% | Slight drop in mechanical properties; improved vascularization; Imposed toxicity | |
| | $Cu^{2+}$ | Co-precipitation | 5 mol% | Low levels of vascularization; lowered cell activity | [26] |
| | $Ba^{2+}$ | Milling and calcination | 1 mol% | Acceptable mechanical properties | [27] |
| | | | 3 mol% | Appropriate mechanical properties | |
| | | | 5 mol% | Prefect mechanical properties | |
| **Monticellite** | $Er^{2+}$ | Co-precipitation | 1 mol% | High degradation rates; improved bioactivity; photoluminescence | [73] |
| | | | 10 mol% | Very high degradation rates; improved bioactivity; risk of toxicity; photoluminescence | |

## 5.1. Fluorine

Fluorine incorporation in bioceramics is well-known with the principal aim of improving stability and retarding the demineralization of enamel in dentistry. Regarding other benefits of this species, $F^-$ contributes to the inhibition of bacterial enzymic reactions, thus keeping a more robust environment in dental areas [74,75]. The addition of $F^-$ also results in the boosted precipitation of fluorapatite which exhibits higher mechanical properties and stability, especially in acidic environments, than hydroxycarbonate apatite [76][77,78].

The addition of $F^-$ ions to Ca-Mg silicates up to a certain level has been proven to enhance the overall behavior of scaffolds. Shahrouzifar et al. [67] doped diopside with $Sr^{2+}$ and $F^-$ through a co-precipitation method and then evaluated the apatite formability and cytocompatibility of the prepared scaffolds. X-ray diffraction (XRD) confirmed the successful incorporation of $F^-$ and $Sr^{2+}$ inside the diopside structure. Using Raman analyses, they revealed that the substitution of F in places of O causes a change in the bond polarization of previous Si-O and present Si-F in favor of the anion, mostly due to the higher electronegativity of F. For instance, Figure 3 depicts the Raman spectra of pure and F-doped diopside samples. Low-intensity peaks at 1049 and 857 $cm^{-1}$ (assigned to the stretch vibrations of non-bridging Si-O bonds) and 665 $cm^{-1}$ (related to the stretch vibrations of bridging Si-O bonds) are a consequence of F incorporation accompanied by a reduction in the number of Si-O bonds. Also, the shift of the 1008 $cm^{-1}$ peak is another evidence of successful F doping, while the peak at 797 $cm^{-1}$ corresponds to the Si-F bond. This addition limited the release of Si but brought about higher bioactivity compared to undoped diopside. The latter outcome is mainly due to the formation of fluorapatite in addition to hydroxycarbonate apatite, explaining the superior apatite-formation ability of the F-doped samples compared to the pure and Sr-doped ones. Electron microscopy validated the higher apatite formation ability of the F-doped than Sr- and co-doped samples. However, MTT results revealed the higher cell viability of the Sr-doped than F-doped samples, where the lowest cell viability was detected for the co-doped samples due to the high affinity of these two dopants and thereby the lower release of the dopants from the scaffolds.

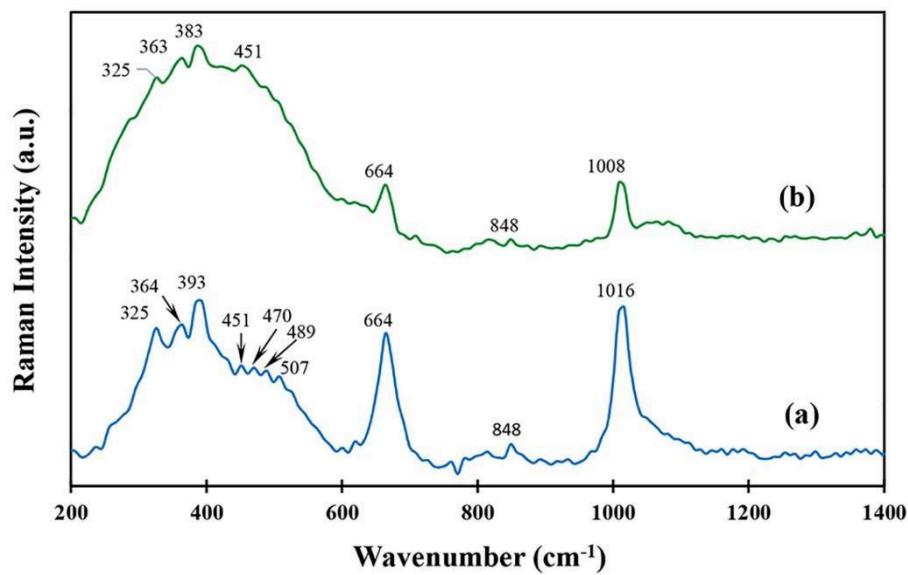

Figure 3. Raman spectra of pure (a) and F-doped (b) diopside scaffolds, adapted with permission from Ref. [67].

In a similar study, Jafari Baghjeghaz and Salahinejad [71] evaluated the structure, apatite-formation ability and ion release of 1 mol% F-doped diopside by using electron microscopy, XRD and inductively coupled plasma-optical emission spectrometry (ICP-OES). The results revealed that the introduction of F$^-$ is attributed to the enlargement of crystallites and the improvement of crystallinity because it lowers the melting and sintering temperatures of diopside, which means the increase in the homologous temperature experienced during sintering. After three days of soaking in the simulated body fluid (SBF), it was found that the undoped specimens show no or little apatite deposition. In contrast, the F-doped one was 85% covered with leaf-like apatite of almost 20 nm in thickness (Figure 4). The ICP-OES results of this study showed a highly decreased amount of P in the simulated body fluid (SBF) for the doped samples compared to the undoped ones. This is due to the improved deposition of apatite on the surface of diopside, confirming the aforementioned results. As a matter of fact, this study emphasizes the effect of fluorine on the improvement of bioactivity.

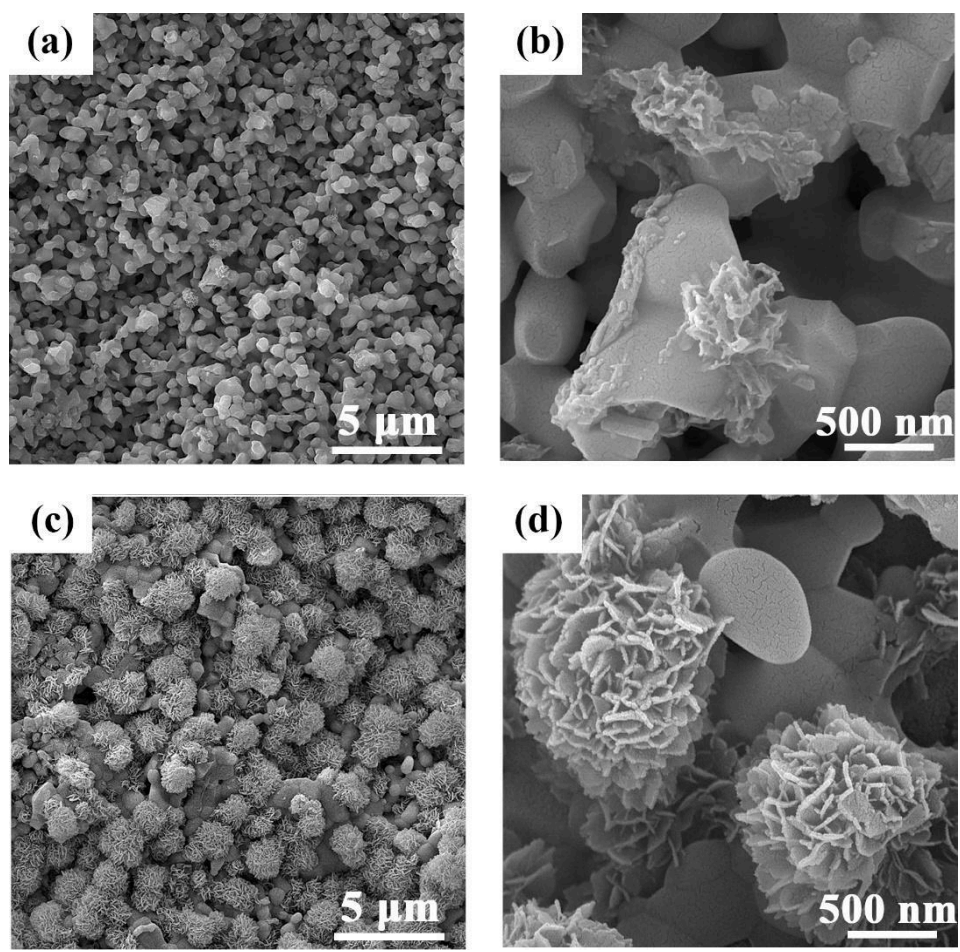

Figure 4. Pure (a, b) and F-doped (c, d) diopside after 3 days of immersion in SBF, reprinted with permission
from Ref. [71].

$F^-$ can moderately enhance pH-related behaviors. $Ca^{2+}$, $Mg^{2+}$ and $Si^+$ releases caused

SBF to become slightly basic, but the incorporation of fluoride has a buffering effect due to replacement with $OH^-$. Regarding 0, 1 and 2 mol% F-doping into diopside, the pH level of SBF first increased until the seventh day of exposure to pure diopside, followed by a fall due to the consumption of $OH^-$ for apatite precipitation. Comparatively, 1 mol% F-doped diopside provided lower pH values over the entire range of the SBF immersion, ensuring the best cell activity due to benefiting from both a buffered physiological medium and effective apatite deposition. However, 2 mol% F doping deprived diopside of apatite precipitation due to the domination of fluorite ($CaF_2$) deposition. Still, it indicated lower pH values than pure diopside due to the buffering effect of $F^-$ [70].

Regarding the incorporation level of fluoride, there is a limit in obtaining desired structures and properties. Keihan and Salahinejad [72] realized that Ca-Mg silicates doped with 0.5 and 1 wt% $F^-$ after sintering form single-phase bredigite and a multi-phase product (consisting of akermanite and diopside), respectively, validated by XRD analyses. In another study, up to 2 wt% $F^-$ formed single-phase diopside, where the higher percentages of fluoride resulted in the formation of fluorite, cuspidine, akermanite and monticellite. Also, the excess usage of $F^-$ caused the slight toxicity of the samples [70]. In an study by Esmati et al. [78], the cytocompatibility and apatite-formation ability of coprecipitated $2SiO_2$-MgO-CaO (the same composition as diopside) bioactive glass scaffolds doped with 0, 0.5, 1, 1.5 and 2 mol% F were evaluated. The best biological behaviors were observed at the optimal level of 1-1.5 % addition. The deterioration beyond 1.5 % is mainly due to two reasons; (1) the deposition of fluorite ($CaF_2$) instead of calcium phosphates, retarding biomineralization [79] and (2) the excessive amount of fluoride release resulting in lower cytocompatibility [80]. It is established that amounts higher than 10 mg $F^-$ per day can be toxic and lead to skeletal fluorosis, weakening of bones and gastrointestinal sicknesses. This is primarily due to reactions between F and acetyl groups in the body, producing fluoroacetic acid that infiltrates the proper completion of the Krebs cycle [81].

Even though $F^-$ is believed to improve bioactivity and biostability, there are a few drawbacks or areas towards improvement for its further use as a dopant. One of its major disadvantages is lowered cell viability compared to dopants like strontium [67]. Co-doping can be an available measure to address this challenge, so that a second dopant compensates for cytotoxicity caused by fluoride.

### 5.2. Strontium

Strontium is a trace species in the human body, which facilitates the differentiation of osteoblastic cells into osteocytes. It is well-established that the improvement of osteogenesis (in two matters of osteoblast reactivity and osteoclast retardation) takes place when Sr is introduced to bone tissues [42,82,83]. It is also worth noting that Sr improves the osteoblast differentiation of mesenchymal stem cells and inhibits the differentiation of osteoclasts [84].

Owing to the higher radius of Sr than Ca and the reduction in O-Si-O bond angles and strength owing to Sr substitution for Ca, slight shifts occur in Fourier-transform infrared (FTIR) spectroscopic vibration and X-ray diffraction (XRD) peaks [57]. For instance, Figure 5 represents the typical shift of the most intensified XRD peak of akermanite after Sr doping from 0 to 0.15 mol%.

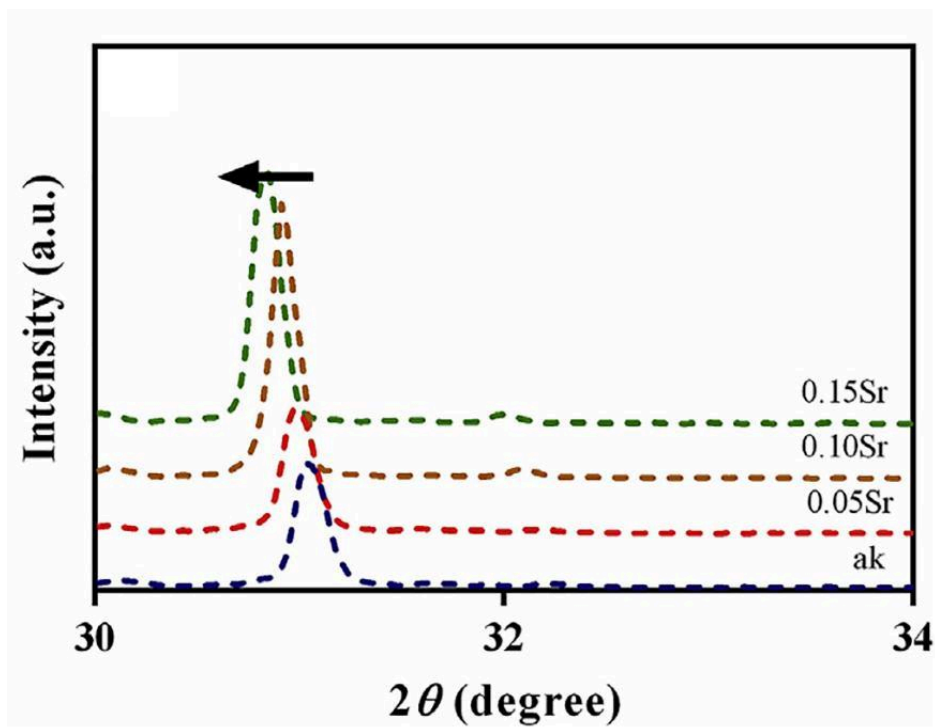

Figure 5. XRD peak shift of akermanite doped with 0.05, 0.10 and 0.15 mol% Sr, reprinted with permission from Ref. [57].

Owing to the size difference of Ca and Sr ions, the replacement of Sr in place of Ca may cause Mg to be released more freely, offering higher apatite-formation ability [85,86]. In contrast, some studies indicate that the same reason would act out as an inhibitory factor in the release of ions having a significant effect on apatite formation, reducing bioactivity [87,88]. These different results are obtained due to two opposing structural characteristics; the increase in the mean distance of bridging bonds (and consequently the decrease in that of non-bridging bonds) and the more ionic character of Sr-O than Ca-O. Typically, Sharouzifar and Salahinejad [67] obtained finer apatite spheres on Sr-doped diopside scaffolds than F-doped ones (1.49 μm compared to 2.29 μm) after 7 days of SBF immersion. In another research [68] on apatite formation and cellular responses by using Fourier-transform infrared spectroscopy (FT-IR), XRD, electron microscopy and energy-dispersive X-ray spectroscopy (EDS), after 7 days in SBF, pure and Sr-doped diopside specimens presented a layer of plate-like and spherical apatites, respectively. The results indicate the higher apatite-formation ability of the Sr-doped samples. In conclusion, it seems to be difficult to state about the effect of Sr doping on bioactivity clearly. Thus, it is best to say that the bioactivity of Ca-Mg silicates mostly depends on their composition, as different compositions displayed different behaviors. Concerning cell activity, Sr-doping can excite mesenchymal stem cell (MSC) adhesion and proliferation on silicates, for instance, diopside (Figure 6). This is related to the controlled release of $Sr^{2+}$, which inhibits surface reactions downgrading protein adsorption [68].

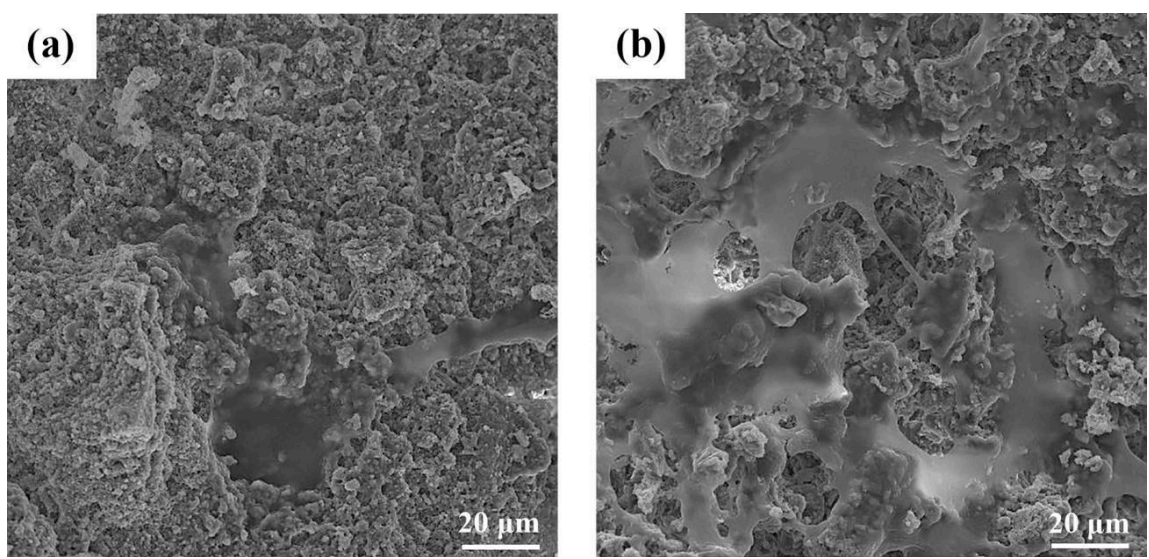

Figure 6. 24-hour incubation of MSCs on pure (a) and Sr-doped (b) diopside, reprinted with permission from Ref. [68].

With regards to mechanical properties, Sr-doping generally improves the fracture toughness, tensile strength and hardness of silicates, for example, akermanite [57], depending on the concentration of strontium. Tensile strength testing results were 12.32, 16.44, 13.59 and 13.82 MPa for the 0, 0.05, 0.10 and 0.15 mol% Sr-doped scaffolds after sintering at 1250° C. As observed, 0.05 mol% was the optimal sample for tensile strength. The same composition was optimal for hardness and fracture toughness with values of almost 4.9 $HV_{0.5}$ and 1.7 $MPa.m^{0.5}$, respectively. This mechanical alteration was attributed to the densification and smaller average grain size of the Sr-doped scaffolds compared to pure ones. Even though the Sr-doped scaffolds were smaller in the grain size than pure akermanite (1.61 μm), there was a rising trend in the grain size by increasing the Sr concentration (from 0.96 μm to 1.59 μm). This is why the 0.05 mol% samples were the optimal ones.

Conclusively, $Sr^{2+}$ is less effective than $F^-$ from the bioactivity viewpoint. But in the matter of mechanical properties, Sr-substituted samples demonstrated better behaviors. In the case of using akermanite, it is suggested that Sr-doping is prior to F and when diopside requires better bioactivity, $F^-$ is a proper dopant.

### 5.3. Copper

Copper is best known for its antibacterial behavior against pathogens in the human body. Another beneficial asset of using $Cu^{2+}$ is its osteoinduction and proliferation of endothelial cells which play a substantial role in the course of angiogenesis and vascularization [89]. Studies also suggest that copper facilitates bone growth without the threat of osteoporosis [90].

He et al. [26] fabricated diopside 3d-printing scaffolds coated with bredigite and akermanite incorporated with $Cu^{2+}$ in place of 5 mol% Ca and compared their ion release, histological and *in vivo* behaviors. The results, in overall, presented high angiogenesis for the Cu-bredigite coatings. The akermanite-Cu coated scaffolds demonstrated a moderate release of Mg over a 7-day immersion in the Tris buffer, while the bredigite-Cu ones were even milder in such behavior. Regarding Cu release, the bredigite-Cu coated scaffolds had the upper hand. Figure 7 visually depicts blood vessels formed in these surface-modified 3D-printed scaffolds, while Figure 8 illustrates the quantitative angiogenesis of the different scaffolds after 6 and 12 weeks of implantation in rabbits. It is revealed that the bredigite-Cu coated scaffolds were the optimal samples with the higher count of blood vessels formed *in vivo* after both 6 and 12 weeks of implantation. In fact, vascular endothelial growth factor (VEGF) is responsible for blood vessel formation. Cu activates hypoxia-inducible factor-1 (HIF-1) through interactions with genes that provoke angiogenesis like VEGF [91]. It is also worth noting that the bioactivity of pure diopside scaffolds was enhanced using bredigite and akermanite coatings since Cu had a slightly destructive effect on bioactivity with lesser cell attachment.

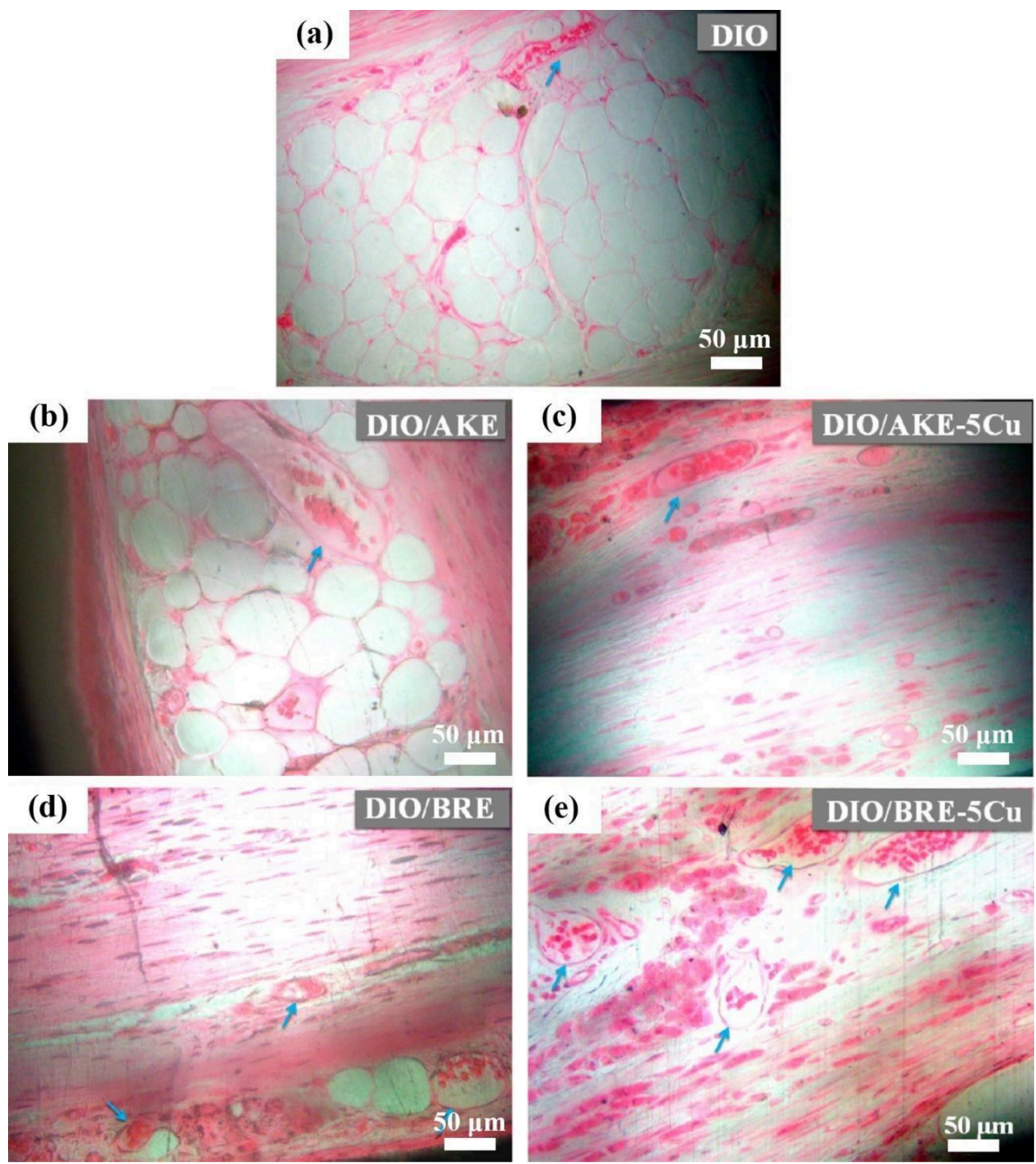

Figure 7. Bare (a), akermanite-coated (b), akermanite-5Cu-coated (c), bredigite-coated (d) and

bredigite-5Cu-coated (e) diopside implants placed in muscle for 12 weeks. Blue arrows indicate vessels grown into pores, reprinted with permission from Ref. [26].

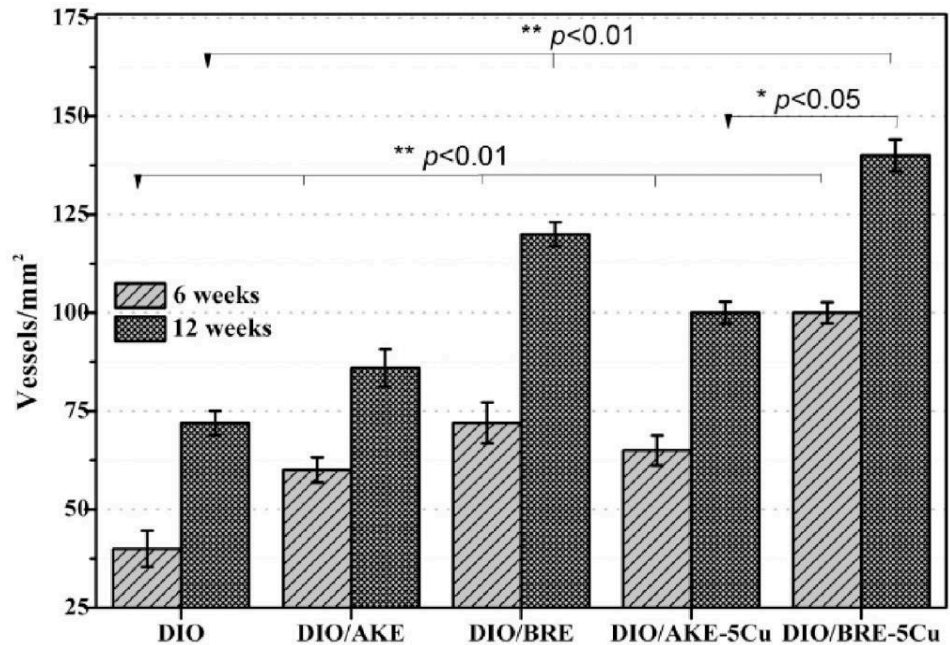

Figure 8. A comparison of vascularization among akermanite, akermanite-5Cu, bredigite and bredigite-5Cu
coatings on 3D-printed diopside scaffolds, reprinted with permission from Ref. [26].

Concerning mechanical aspects, Cu improves the hardness, fracture toughness and strength of silicates. 0, 2.5, 3.5 and 4.5 mol% Cu-doped akermanite scaffolds were fabricated through a high-energy mechanical activation method followed by sintering [58]. The scaffolds with the higher $Cu^{2+}$ contents became denser and their average grain size increased [92]. However, Cu acted as a grain boundary strengthening agent; thus, transgranular fracture was observed as the dominating fracture mode. A rising trend of hardness (from about 0.8 GPa to 6.2 GPa, Figure 9) and fracture toughness (from almost 0.7 $MPa.m^{0.5}$ to 2.4 $MPa.m^{0.5}$) was achieved by increasing the Cu level. Despite these beneficial effects, there was a slight drawback in the mechanical properties from 3.5 mol% to 4.5 mol% Cu, which is maintly attributed to the creation of a glassy-kind phase.

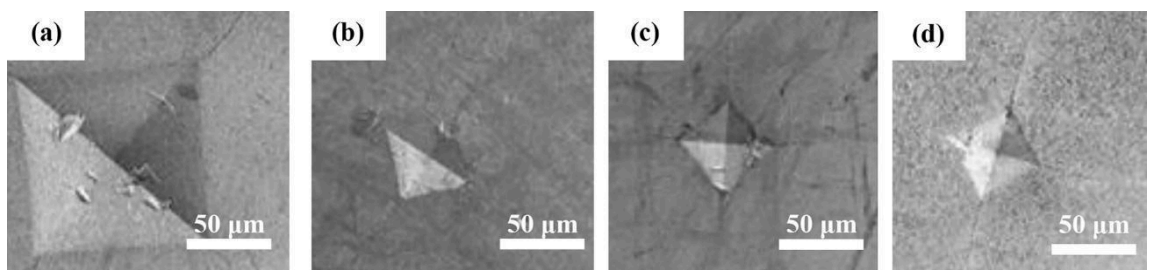

Figure 9. Vickers hardness indentation on 0 (a), 2.5 (b), 3.5 (c) and 4.5 (d) mol% Cu-doped akermanite, reprinted with permission from Ref. [58].

To put it briefly, the most efficient role of copper as a dopant is its antibacterial and angiogenesis. It is not much help in the coordination of use for bioactivity, as it damages cells at higher concentrations. However, it holds an appropriate potential to enhance the mechanical properties of Ca-Mg silicates. Lack of research is sensed on a more profound understanding of copper effects on Ca-Mg silicate, demanding more investigations.

### 5.4. Europium

The use of REEs in bioceramics is chiefly dedicated to imaging abilities they add to products. The preference for photoluminescence, which can occur due to the addition of $Eu^{2+}$, to other imaging activation methods is due to its ability to conduct easy diagnosis in a high rate of data collection and visualization of biological mechanisms and implanted substitutes in real time [73]. $Eu^{2+}$-doping into bioceramics has been proven to emit IR waves from mice tissues, holding a prior UV light treatment.

The bioactivity of Eu-doped Mg-Ca silicates is improved by increasing the Eu content. Other studies on bioactive glasses have also proposed the same results [29,93,94]. The addition of Eu to monticellite via slurry mixing from boron wastes and $Si_3N_4$ has been proven to be successful in the matters of bioactivity through improving the surface area via increasing the surface roughness [73]. It is well-established that surfaces of 20 μm in roughness can improve cell attachment, whereas larger and smaller pore sizes in the same order may result in lower cell attachment [95,96]. Sintered monticellite scaffolds containing 1 and 10 wt% Eu were studied in terms of apatite-formation ability [73]. The increase in the sintering temperature from 900 °C to 1000 °C resulted in the creation of an amorphous phase and the reduction in the pore size and number, confirmed by electron microscopy and EDS. Intensification of the Eu content resulted in a narrow pore size distribution and the reduction of the mean pore size (from 30 μm to 15-20 μm) while having a higher number. Besides, some amorphous regions can be also overseen due to the fast kinetics of sintering in the presence of $Eu_2O_3$ as a sintering agent. This decrease in the pore size accompanied by a higher number of pores gave rise to higher roughness, knowing that a greater dissolution rate of ceramics results from the increase of roughness and surface area [97]. Thus, the increment of the surface roughness resulted in the formation of calcium phosphate on the samples with the higher $Eu^{2+}$ content, where the rise in the temperature from 900 °C to 1000 °C caused the formation of smoother but denser tetracalcium phosphate ($Ca_4O(PO_4)_2$).

To the authors' knowledge, there is a lack of evidences supporting the mechanical properties of medical Eu-doped Ca-Mg silicates. The authors cannot even confirm that there are any studies regarding the mechanical properties of Eu-substituted bioactive glasses. In conclusion, bioactivity was generally enhanced after the addition of europium.

## *5.5. Barium*

Barium is not an essential species in the human body metabolisms, but it contributes to bone formation via enhancing the apatite formation of Ca-Mg silicates. Moreover, barium is known as a strengthening agent. Anti-inflammatory properties of barium incorporated in 45S5 bioglass fabricated by sol-gel method have been studied by Majumdar et al. [98]. Human exposure to 3-4 gr Ba per day is considered toxic [93].

Myat-Htun et al. [27] prepared Ba-doped (1, 3 and 5 mol%) akermanite scaffolds and evaluated their structure, mechanical and apatite formation behaviors by using XRD, Vickers' hardness test, electron microscopy and EDS. The results, in general, indicated improvements in mechanical properties and bioactivity. Ba worked as a sintering agent, improving the relative density from 62.67% to 94.25% (for pure to 5 mol% Ba-doping) after sintering at 1200 °C through liquid-phase sintering, as the same trend was observed by Ref. [99]. This would bring about high mechanical properties, for instance, up to two-to-threefold tensile strength. Moreover, the hardness and fracture toughness of the Ba-doped scaffolds were improved, valuing from about 0.9 GPa to 5.1 GPa and from almost 0.7 MPa.m$^{0.5}$ to 1.4 MPa.m$^{0.5}$, respectively. As expected, the pure samples had the lowest values in all categories, whereas 5 mol% Ba was the best acting one.

Pure akermanite is known to produce apatite in SBF after 21 days of immersion with a spherical morphology [49,57]. The addition of Ba and the rise in its content lead to the formation of worm-like apatites on the samples in the same period. Besides, it was observed that increasing the barium content takes part in better apatite-formation ability as the whole surface of the 5 mol% Ba-doped akermanite scaffolds was covered with apatite. The ionic character of Ba-O is slightly higher than Ca-O on account of the comparable electronegativity of Ba (0.9) to that of Ca (1) [100]. The lower covalent character of this bond leads to faster release kinetics; thus, an improvement in apatite precipitation and accordingly bioactivity take place [27].

Although there is insufficient data to rely on, it is concluded that Ba improves both mechanical and biological behaviors to extreme values. Besides, it seems that until there is no cytocompatibility issues, Ba can be used for higher contents.

### 5.6. Cerium

Cerium as an REE, holding a close electronegativity and size in the ionic state ($Ce^{3+}$) to $Ca^{2+}$ [100], presents inflammatory [101] and antibacterial [102] behaviors. It indicates potential areas that can participate in Ca-Mg silicates as a doping agent. Su et al. [28] revealed that diopside doped with $Ce^{3+}$ brings about better mineralization and strength up to a certain content of cerium. $Ce^{3+}$ caused a slight change in the structure of silicates, so that the increment of cerium from 0.25 mol% to 1 mol% altered the whole composition of diopside to $CeO_2$ and $MgSiO_3$. This phase change resulted in a lower dissolution rate; thus, apatite formability was significantly reduced after an improvement to a doping level of 0.25 mol% $Ce^{3+}$. A constant rise in pH for this level of incorporation in comparison to slight and no increase is evidence for the higher dissolution rate of the 0.25 mol% samples than the 0.5 and 1 mol% ones, respectively.

The compressive strength of the Ce-doped samples was enhanced from 0 to 0.25 mol% Ce, then reduced at 0.50 mol% Ce and at last increased to a superior level at 1.00 mol% Ce. This trend might be a result of structural changes occurring with the addition of Ce. When the Ce content reached 0.25 mol%, the formation of the high-strength $CeO_2$ phase in the major diopside matrix improves overall strength (near 130 MPa). At 0.50 mol% Ce, the low-crystallinity composition of $CeO_2$ and $MgSiO_3$ declined compressive strength to approximately 50 MPa. The increase of the Ce content to 0.75 or 1.00 mol% caused an upsurge in strength on account of the high crystallinity of the aforementioned phases [28].

From the mechanical viewpoint, it seems that the alteration of the base ceramic to phases like $CeO_2$ and $MgSiO_3$ may first decrease and then increase the strength. Regarding the biological effect of Ce-doping, it was observed that before the formation of other phases, appropriate properties could be obtained, whereas at higher Ce contents lower degradability reduces bioactivity.

### 5.7. Alkali elements (lithium, sodium and potassium)

Alkali cations of $Li^+$, $Na^+$ and $K^+$ have been used by Rahmani and Salahinejad [69] as dopants in diopside. These cations are trace elements in the human body, contributing to different metabolisms. Lithium improves bioactivity [103] and provides antibacterial effects toward certain bacterial strains [104]. Sodium and potassium are mostly strengthening factors rather than biological agents [105]. Sodium is a major participant of bioactive glass systems due to its network modifying properties [106]. The addition of these monovalent cations to diopside in the content of 2 mol% (to replace 1 mol% $Ca^{2+}$) caused lower crystallinity. Moreover, distortion caused by $K^+$ was more than $Na^+$ as $K^+$ is larger in size than $Na^+$ [100]. Nonetheless, lithium interstitial accommodation in the silicate network gave rise to a higher crystallinity than the other two doped diopside scaffolds.

Regarding biological behaviors, K-doped scaffolds formed spherical colonies of apatite after 7 days of SBF immersion. Full coverage of apatite with uniform nanometric plates was observed on the surface of Li-doped. Na-doped samples revealed lower apatite formation than Li-doped ones, but more than pure diopside. According to Figure 10, the MTT assay indicates that MG-63 cell proliferation is improved significantly using $K^+$ and $Na^+$. Additionally, $Li^+$ was the last-ranked in cell viability among the doped samples. It is worth noting that all of the doped samples were superior to undoped diopside [69]. In conclusion, potassium was the optimal dopant in both cases of bioactivity and MG-63 cell reactivity in comparison to the other doping agents.

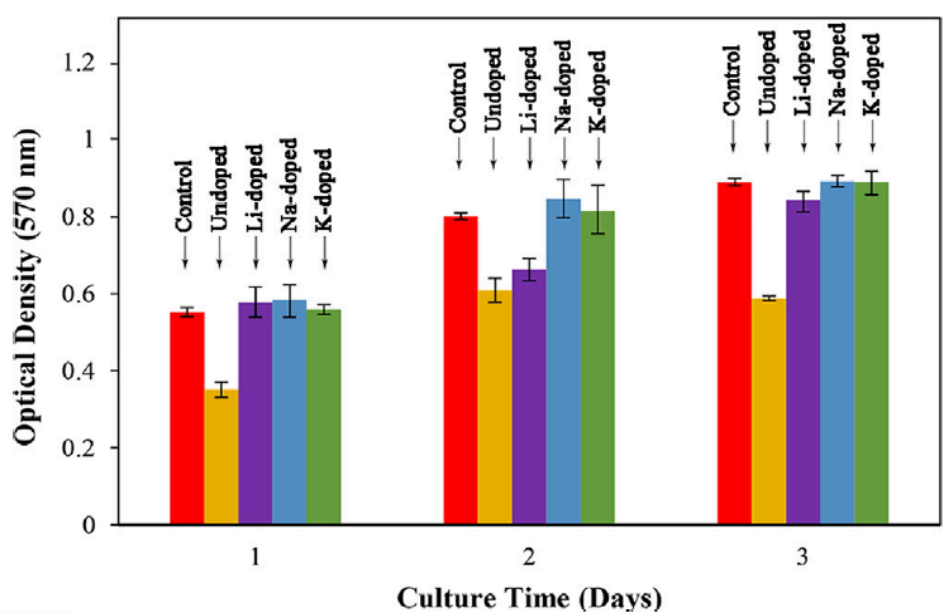

Figure 10. MTT assay results of MG-63 cell cultures on Li-, Na- and K-doped diopside scaffolds, reprinted with permission from Ref. [69].

## 6. Conclusions and future outlooks

Ion-doping is deduced as an easy, yet effective way to enhance the mechanical and biological behaviors of Ca-Mg silicates used in biomedicine with some drawbacks, as summarized in Table 5. It is worth noting that the conclusions have been made with the caution of dopant concentrations. That is, excessive amounts of any ions in the human body can result in electrolytic disorders (e.g., in the case of Na which results in brain disfunction, as called hypernatremia [107]).

Table 5. A summary of the contributions of dopants used in Ca-Mg silicates

| Dopant | Mechanical properties | Apatite formability | Osteoinduction / Osteoconduction | Angiogenesis | Biocompatibility |
|---|---|---|---|---|---|
| **Fluorine** | Mechanical properties are rather unchanged* | Improves apatite-formation ability by the formation of fluorapatite | Not reported | Not reported | Introduces a level of toxicity after reaching its critical concentrations (e.g., 1.5 mol% F⁻ for diopside); Buffers SBF in concentrations near 1 mol% for diopside resulting in better cytocompatibility; Promotes antibacterial behaviors* |
| **Strontium** | Improves strength and fracture toughness at optimal limits (e.g., 0.05 mol% for akermanite) | Improves apatite-formation ability | Improves bone regeneration* (osteoconduction); Improves cell differentiation (osteoinduction) | Improves vascularization* | Overexposure may result in bone abnormalizations in young ages* |
| **Copper** | Improves mechanical properties considerably up to a certain content (e.g., 3.5 mol% for akermanite); Slightly worsens mechanical properties at elevated contents due to the introduction of glassy phases | Not reported | Reduces bioactivity at high levels by lowering cell attachment | Improves vascularization considerably | Cytotoxicity and lessened cell attachment at high contents; Antibacterial behavior at optimal concentrations |
| **Europium** | Higher hardness can be obtained* | Slightly improves apatite-formation ability | Not reported | Not reported | Cytotoxicity at high contents |
| **Barium** | Introduces great mechanical properties | Improves apatite-formation ability by the formation of worm-like apatite | Not reported | Not reported | No biocompatibility issues have been observed |
| **Cerium** | Unstable improvement of mechanical properties due to partial or complete phase transformations | Improves apatite-formation ability in an optimal concentration (e.g., 0.25 mol% for diopside) due to higher dissolution rates; Reduces bioactivity at concentrations at which $CeO_2$ and $MgSiO_3$ are present | Enhances the mineralization of bone (osteoconduction) | Not reported | Cytotoxicity at high contents* |
| **Lithium** | Induces better mechanical strength* | Provides apatite-formation ability despite | Improves cell proliferation (osteoconduction) | Not reported | Lowers cell viability at high concentrations; Offers bactericidal behaviors |
|  |  | lowering apatite deposition |  |  |  |
| **Sodium** | Offers improved mechanical properties | Insufficient apatite-formation ability | Not reported | Not reported | Not reported |
| **Potassium** | Not reported | Improves apatite-formation ability | Improves cell activity (osteoconduction) | Not reported | No cytotoxicity issues have been reported |

*Reported in studies other than Ca-Mg silicates

Regarding the characteristics listed in Table 3 and the characteristics of diopside, akermanite, bredigite and monticellite, it is possible to draw out the best choice of dopants for each Ca-Mg silicate (Table 6). However, there are still some possibilities for further investigations. Looking at Table 6, it is clear that some of the dopants which might possess a positive influence on the properties of Ca-Mg silicates have not been still studied. Moreover, the introduction of elements like Ag (antibacterial behavior [108]), Fe (metabolic reactions and osteoconductive properties [109,110]), B (angiogenesis properties [108]), Zn (osteoconductive properties [111] and angiogenesis [112] ), Cr (degradation, angiogenesis and antibacterial properties [113]), Co (angiogenesis and bactericidal behaviors [114]), etc. are some other options. To our knowledge, no medical products of this ceramic system exist in the market, which brings abouts an outstanding potential for further research and development.

Table 6. Suitable dopants for each Ca-Mg silicate (Y and N mean reported and unreported, respectively)

| Silicate name / Dopant | Diopside | Akermanite | Bredigite | Monticellite |
|---|---|---|---|---|
| F | Y [67,70,71] | N | N [46] | Y |
| Sr | Y [67,68] | Y [57] | Y | Y |
| Cu | N [26] | Y [26,58] | Y with biocompatibility caution [26] | N |
| Eu | N | N | N | Y with biocompatibility caution [73] |

| | | | Y with biocompatibility caution | |
|---|---|---|---|---|
| **Ce** | N [28] | Y | | N |
| **Ba** | Y | Y [27] | Y | Y |
| **Li** | Y [69] | N | N | Y |
| **Na** | N [69] | Y | Y | N |
| **K** | Y [69] | N | N | Y |
| **Ag** | N | N | N | N |
| **Fe** | N | N | N | N |
| **B** | N | N | N | N |
| **Zn** | N | N | N | N |
| **Cr** | N | N | N | N |
| **Co** | N | N | N | N |